\documentclass[12pt,a4wide]{article}
\usepackage{graphicx}

\begin{document}

\newcommand{\lsim}{\mbox{\raisebox{-.9ex}{~$\stackrel{\mbox{$<$}}{\sim}$~}}}
\newcommand{\gsim}{\mbox{\raisebox{-.9ex}{~$\stackrel{\mbox{$>$}}{\sim}$~}}}

\begin{center}
{\Large\bf Is the Big Rip unreachable?}
\end{center}

\medskip

\noindent
\begin{center}
{\large Konstantinos Dimopoulos$^*$}
\end{center}

\medskip

\begin{center}
\noindent
$^*${\small\em Consortium for Fundamental Physics, Physics Department,}\\
{\small\em Lancaster University, Lancaster LA1 4YB, UK}\\
{\small e-mail: {\tt k.dimopoulos1@lancaster.ac.uk},}\\
\end{center}

\medskip

\begin{abstract}
I investigate the repercussions of particle production when the Universe is 
dominated by a hypothetical phantom substance. I show that backreaction
due to particle production prevents the density from shooting to infinity at a
Big Rip, but instead forces it to stabilise at a large constant value. 
Afterwards there is a period of de-Sitter inflation. I speculate that this 
might lead to a cyclic Universe.
\end{abstract}

A phantom substance is defined as a fluid which violates the null energy 
condition because its pressure is \mbox{$p<-\rho$}, where $\rho$ is its 
density. Such a hypothetical substance is also called exotic matter, and it is
necessary for keeping a wormhole traversable. In cosmology, if the Universe
content is dominated by phantom density, then the latter increases and at some 
finite time $t_{\rm max}$ it becomes infinite, in a singularity called the Big 
Rip. This is because the Universe undergoes super-inflation where the
accelerated expansion gives rise to an event horizon, whose dimensions shrink 
to zero at $t_{\rm max}$, ripping apart all structures, from galaxies to atoms.

But is it so? So far, most of the studies of the dynamics of the Universe when 
undergoing accelerated expansion have ignored particle production due to the 
event horizon, because it is usually negligible, in inflation for example. 
In this work, it is argued that particle production
cannot be ignored when the accelerated expansion is driven by phantom density. 
Backreaction due to particle production renders the Big Rip singularity 
unreachable.

We work in Einstein gravity only. In the following, we use natural units, where 
$c=\hbar=1$ and Newton's gravitational constant is \mbox{$8\pi G=m_P^{-2}$}, 
with \mbox{$m_P=2.43\times 10^{18}\,$GeV} being the reduced Planck mass. For 
simplicity, isotropy and spatial flatness is assumed throughout.

The existence of an event horizon
during accelerated expansion results in particle production of all light
(meaning with mass smaller than the Hubble rate \mbox{$m<H$}) non-conformally
invariant fields (e.g. a light scalar field, such as the inflaton). During
quasi-de~Sitter inflation, particle production results in the gravitational 
generation of density of the order of thermal density with temperature the 
Hawking temperature of de~Sitter space \mbox{$T=H/2\pi$}
\cite{parker,brand,larry,habib,EJChun,haro,rigopoulos}\footnote{%
Note that, particle production is a non-equilibrium effect and in an isotropic 
universe its classical counterpart is bulk viscosity \cite{barrow0}.}
\begin{equation}
\rho_{\rm gr}=q\frac{\pi^2}{30}g_*^{\rm gr}\bigg(\frac{H}{2\pi}\bigg)^4\,,
\label{rhogr}
\end{equation}
where $g_*^{\rm gr}$ is the number of effective relativistic degrees of freedom
which undergo particle production (i.e. are light and non-conformally
invariant) and \mbox{$q\sim 1$} is some constant factor due to the fact
that the resulting $\rho_{\rm gr}$ is not actually thermal.

Thus, the Friedmann equation obtains an additional term of the form
\begin{equation}
H^2=\frac{8\pi G}{3}\rho+CH^4\,,
\label{modfried}
  \end{equation}
where $\rho$ is the density of the substance which causes the accelerated
expansion and, in view of Eq.~(\ref{rhogr}), we have
\begin{equation}
  CH^4=\frac{8\pi G}{3}\rho_{\rm gr}
  \;\Rightarrow\;
C=\frac{qG}{180\pi}\,g_*^{\rm gr}\,.
\label{C}
\end{equation}

Strictly speaking, the above are true for \mbox{$\rho,H\simeq\,$constant}, which
results in quasi-de~Sitter inflation. In this case, the $CH^4$ contribution in
the Friedmann equation (\ref{modfried}) is negligible and is usually ignored 
(but see Ref.~\cite{barrow}). However, one expects the same phenomenon to take 
place in any kind of accelerated expansion, since gravitational particle 
production occurs whenever there is an event horizon (as with black hole 
radiation, for example). Thus, qualitatively, the same source term $CH^4$ would 
appear in the Friedmann equation as $H$ sets the scale of the accelerated 
expansion and \mbox{$C\propto g_*^{\rm gr}$} as in Eq.~(\ref{C}).\footnote{%
In general, in accelerated expansion the Hawking temperature is 
\mbox{$T=\frac12|3w+1|(H/2\pi)$} \cite{arty}. The numerical factor 
$\frac12|3w+1|$ can be incorporated into $q$ in Eq.~(\ref{rhogr}).} 
This term would become important if the cause of accelerated expansion is a 
phantom substance.\footnote{In principle, other terms involving the derivatives 
of the Hubble rate, such as $H\ddot H$, $H^2\dot H$ or $(\dot H)^2$ may also be
important, additionally to the $H^4$ term considered here. Such terms are 
negligible in quasi-de Sitter inflation because \mbox{$|\dot H|\ll H^2$}, but 
this would not necessarily be so for a phantom substance. However, in 
Ref.~\cite{paulnew} it
is shown that the density of such derivative terms is proportional to $\alpha$, 
where the latter is the coefficient of $R^2$ in a modified gravity Lagrangian 
\mbox{${\cal L}=\frac12 m_P^2 R+\alpha R^2$}. In this work we consider Einstein 
gravity only, which means \mbox{$\alpha=0$} and these terms are absent.}

If $\rho$ is the density of a phantom substance then the Friedmann
equation (ignoring the extra term $CH^4$, because it can be negligible at first)
results in \mbox{$\dot\rho>0$} and \mbox{$\dot H>0$}. This results in
super-inflation which leads to the Big Rip singularity when
\mbox{$\rho\rightarrow\infty$} in finite time. However, the growth of $H$ means
that the $CH^4$ in the Friedmann equation (\ref{modfried}) will eventually
become important, because it increases faster than the $H^2$ term. 
This may halt the growth of $\rho$ and prevent the Big Rip from happening
\cite{paulold,barrow+}.

The solution to Eq.~(\ref{modfried}) is
\begin{equation}
  H^2=\frac{1}{2C}\bigg(1\pm\sqrt{1-\frac{32\pi GC}{3}\rho}\;\bigg)\,.
\label{Hsolu}
\end{equation}
For small density, the above equation results in either 
\mbox{$H^2=1/C=\,$constant} or
\mbox{$H^2=(8\pi G/3)\rho$}, which is the usual Friedmann equation. The latter 
solution corresponds to the negative sign in the brackets. However,
the above also shows that there is a {\em maximum} possible value of $\rho$,
which is \mbox{$\rho_{\rm max}=3/32\pi GC$}, in which case there is a maximum
value of the Hubble rate \mbox{$H_{\rm max}=1/\sqrt{2C}$}. This means that the 
Big Rip is unreachable, prevented by the backreaction of the gravitational 
particle production.\footnote{A similar result was obtained in 
Ref.~\cite{odintsov1}, where the effect of the conformal anomaly
was taken into account, which can also produce a contribution 
\mbox{$\delta\rho\propto H^4$}.} 

Using Eq.~(\ref{C}), we can obtain an estimate of the maximum density
\begin{equation}
  \rho_{\rm max}=\frac{3}{32\pi GC}=\frac{135}{8qG^2g_*^{\rm gr}}
=\frac{3H_{\rm max}^2}{16\pi G}
  \;\Rightarrow\;
  \rho_{\rm max}^{1/4}=\sqrt{6\pi}\bigg(\frac{30}{qg_*^{\rm gr}}\bigg)^{1/4}m_P\,.
\label{rhomax}
\end{equation}
Thus, $\rho_{\rm max}^{1/4}$ can be smaller than the Planck scale only when
$g_*^{\rm gr}$ is very large. Considering string theory, we can reduce 
$\rho_{\rm max}^{1/4}$ to the string scale $\sim 10^{17}\,$GeV by considering
\mbox{$g_*^{\rm gr}\sim 10^5$} or so. We assume that 
\mbox{$\rho_{\rm max}^{1/4}<m_P$} such that quantum gravity considerations can be 
ignored.\footnote{In a quantum gravity setup, the $H^4$ correction in the 
Friedmann equation~(\ref{modfried}) might be the leading order to terms 
proportional to $H^6$ or $H^8$ for example. Such terms would grow even faster 
than the $H^4$. However, we expect such terms to be Planck suppressed such that 
they always remain subdominant because \mbox{$H^{4+n}/m_P^n<H^4$} for all 
\mbox{$n\geq 1$} since \mbox{$H\leq H_{\rm max}\ll m_P$} because 
\mbox{$\rho_{\rm max}^{1/4}<m_P$}.}

Exactly how the density evolves can be revealed by the study of the continuity
equation
\begin{equation}
\dot\rho+3(1+w)H\rho=0
\,,
\label{cont}
\end{equation}
where $w=p/\rho<-1$ is the barotropic parameter of the phantom substance. 
For simplicity, we consider \mbox{$w=\,$constant}.
Using Eq.~(\ref{Hsolu}) with the negative sign in the brackets, 
Eq.~(\ref{cont}) can be analytically solved to give
\begin{equation}
\frac{1}{\sqrt 2}\ln\left(\frac{\sqrt 2+\sqrt{2C} H}{\sqrt 2-\sqrt{2C} H}
\frac{\sqrt 2-1}{\sqrt 2+1}\right)-2\left(1-\frac{1}{\sqrt{2C}H}\right)=
\frac{3|1+w|}{\sqrt{2C}}(t_{\rm max}-t)\,,
\label{soluH}
\end{equation}
where $t_{\rm max}$ corresponds to the time when \mbox{$\rho=\rho_{\rm max}$}.
Note that Eq.~(\ref{Hsolu}) (with the negative sign in the brackets) implies 
that \mbox{$\sqrt{2C}\,H=(1-\sqrt{1-u})^{1/2}$},
where we have defined \mbox{$u\equiv\rho/\rho_{\rm max}$}, i.e. 
\mbox{$u(t_{\rm max})=1$}. 

When \mbox{$H\ll H_{\rm max}=1/\sqrt{2C}$}, the last term in the left-hand-side of
the above dominates and we find \mbox{$H^{-1}=\frac32|1+w|(t_{\rm max}-t)$} as 
with standard phantom dark energy, only the time $t_{\rm max}$ does not denote the
Big Rip, i.e. \mbox{$\rho(t_{\rm max})\not\rightarrow\infty$}, but instead we have
 \mbox{$\rho(t_{\rm max})=\rho_{\rm max}$} which is finite and given by 
Eq.~(\ref{rhomax}).

The solution in Eq.~(\ref{soluH}) only applies for \mbox{$t\leq t_{\rm max}$}. At 
$t_{\rm max}$ the solution becomes zero.%
\footnote{In view of Eqs.~(\ref{rhogr}) and (\ref{rhomax}), we have
\mbox{$
\rho_{\rm gr}\leq\rho
\;\Leftrightarrow\;8\pi G\rho/3\geq CH^4 
$}.
This means that we cannot connect with the positive branch of $H^2$ in 
Eq.~(\ref{Hsolu}), which ranges between $1/\sqrt{2C}$ and $1/\sqrt C$, because 
this requires \mbox{$8\pi G/3\rho<CH^4$}.}
We can investigate what happens when \mbox{$t>t_{\rm max}$} by considering the 
continuity equation~(\ref{cont}), which gives
\begin{equation}
\dot u=3|1+w|Hu\;\Rightarrow\; \dot u_{\rm max}=\frac{3|1+w|}{\sqrt{2C}}>0\,,
\label{contu}
\end{equation}
where \mbox{$\dot u_{\rm max}\equiv\dot u(t_{\rm max})$} and
we used that \mbox{$H_{\rm max}=1/\sqrt{2C}$} and \mbox{$u(t_{\rm max})=1$}.
Thus, there is a tendency for the density to become larger than $\rho_{\rm max}$,
which, however, is forbidden by Eq.~(\ref{Hsolu}). This is really because
Eq.~(\ref{modfried}) is not valid any more. Indeed, for 
\mbox{$\rho>\rho_{\rm max}$}, the density of the gravitationally produced 
particles would be larger than the phantom density itself, which cannot happen
(where is the energy coming from?). 

This suggests that the continuity equation needs augmenting, since the removal 
of energy by gravitational particle production is not considered in 
Eq.~(\ref{contu}); it is implicitly assumed negligible. To take this into 
account we introduce a negative source term on the right-hand-side of 
Eq.~(\ref{cont}), which becomes (see also Ref.~\cite{tommi+malcolm})
\begin{equation}
\dot\rho-3|1+w|H\rho=-3|1+w|H\rho_{\rm gr}=-\frac{9|1+w|C}{8\pi G}H^5\,.
\label{contH5}
\end{equation}
The form of this term is given by \mbox{$\delta\rho/\delta t$}, where
\mbox{$\delta\rho=-\rho_{\rm gr}\propto H^4$} as in Eq.~(\ref{rhogr}) and
\mbox{$\delta t\sim H^{-1}$} is the Hubble time. This is because the
relativistic particles produced gravitationally in a Hubble time are diluted
by the accelerated expansion and replenished by the particles produced in the
following Hubble time, meaning that \mbox{$\delta\rho\propto H^4$} is produced
gravitationally per Hubble time \mbox{$\delta t\sim H^{-1}$}. The precise value 
is \mbox{$\delta t=H^{-1}/3|1+w|$}, which results in \mbox{$\dot u_{\rm max}=0$}.

Combining Eqs.~(\ref{Hsolu}) and (\ref{contH5}) we may write the continuity 
equation as
\begin{equation}
\dot u=6|1+w|H_{\rm max}\sqrt{1-u}(1-\sqrt{1-u})^{3/2}\,,
\label{augcont}
\end{equation}
where \mbox{$H_{\rm max}=1/\sqrt{2C}$}. From the above it is evident that 
\mbox{$\dot u_{\rm max}=0$} as expected. Also note that, in the limit 
\mbox{$u\ll 1$} the above becomes 
\begin{equation}
\dot u\simeq\mbox{$\frac32$}|1+w|\sqrt 2H_{\rm max}u^{3/2}\simeq 3|1+w|Hu\,,
\end{equation}
which is the usual continuity equation (cf. Eq.~(\ref{contu})) and we 
considered that 
\mbox{$H=H_{\rm max}(1-\sqrt{1-u})^{1/2}\simeq H_{\rm max}u^{1/2}/\sqrt 2$} in this 
limit. Note that in the limit \mbox{$u\ll 1$} we recover the Friedmann 
equation as expected because 
\mbox{$H^2\simeq\frac12 H_{\rm max}^2u=8\pi G\rho/3$}, where
\mbox{$H_{\rm max}=1/\sqrt{2C}$} and $\rho_{\rm max}$ is given by 
Eq.~(\ref{rhomax}).

The solution to Eq.~(\ref{augcont}) is
\begin{equation}
\frac{\rho}{\rho_{\rm max}}=u=1-\left\{1-\left[\mbox{$\frac32$}|1+w|H_{\rm max}
(t_{\rm max}-t)+1\right]^{-2}\right\}^2.
\label{solu}
\end{equation}
If we consider the limit \mbox{$t\ll t_{\rm max}$} and the condition 
\mbox{$H_{\rm max}t_{\rm max}\gg 1$} we find that the above asymptotes to the value
\mbox{$\rho\rightarrow 1/6(1+w)^2t_{\rm max}^2$}. This is the same value one 
obtains in the limit \mbox{$t\ll t_{\rm max}$} for standard phantom dark 
energy, for which \mbox{$\rho=1/6(1+w)^2(t_{\rm max}-t)^2$}. 

\begin{figure}[h]
\centering
\vspace{-8cm}
\includegraphics[width=1\linewidth]{%
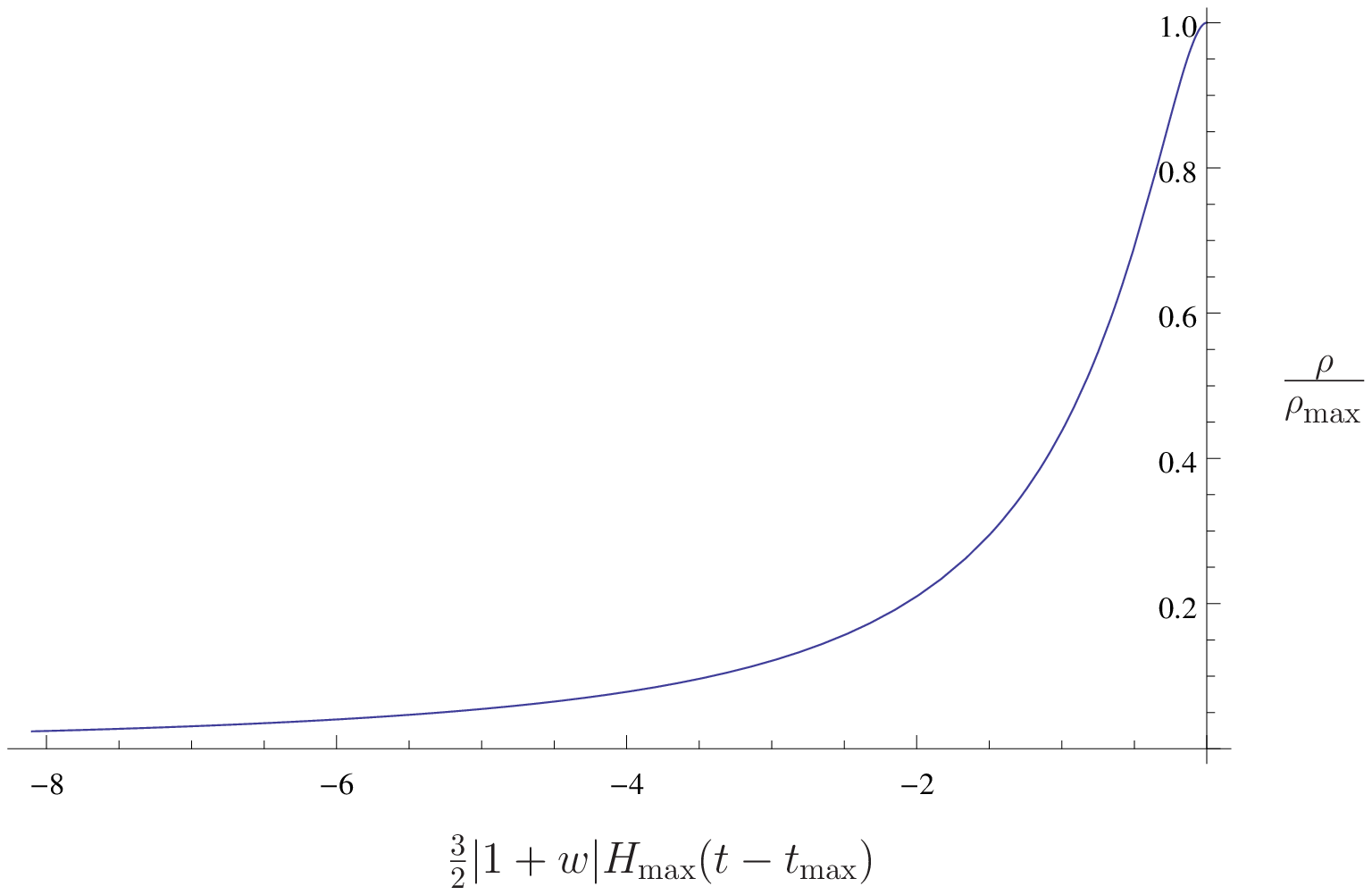}
\vspace{-5.5cm}
\caption{%
Evolution of \mbox{$u=\rho/\rho_{\rm max}$} as time approaches $t_{\rm max}$. 
The Big Rip is avoided and the density approaches a constant value 
$\rho_{\rm max}$.}
\label{phantomfig}
\end{figure}

The evolution of the density $\rho$ until the time $t_{\rm max}$ is shown in 
Fig.~\ref{phantomfig}. As we approach $t_{\rm max}$, instead of shooting to 
infinity, the backreaction due to the particle production forces the growth
of $\rho$ to be halted and the latter gently reaches $\rho_{\rm max}$ at 
\mbox{$t=t_{\rm max}$}. What happens afterwards? Well, the constant values 
\mbox{$\rho=\rho_{\rm max}$} (given by Eq.~(\ref{rhomax})) and 
\mbox{$H=H_{\rm max}=1/\sqrt{2C}$} are solutions to Eqs.~(\ref{modfried}) and
(\ref{augcont}) so that the density and the Hubble parameter remain constant.
Thus, the Universe undergoes de-Sitter inflation, even though it is filled at 
equal parts by a phantom substance with \mbox{$w<-1$} and radiation due to 
particle production which is constantly diluted and replenished so that it has
a constant net density \mbox{$\rho_{\rm gr}\sim H_{\rm max}^4$}.

\begin{figure}[h]
\centering
\includegraphics[width=1\linewidth]{%
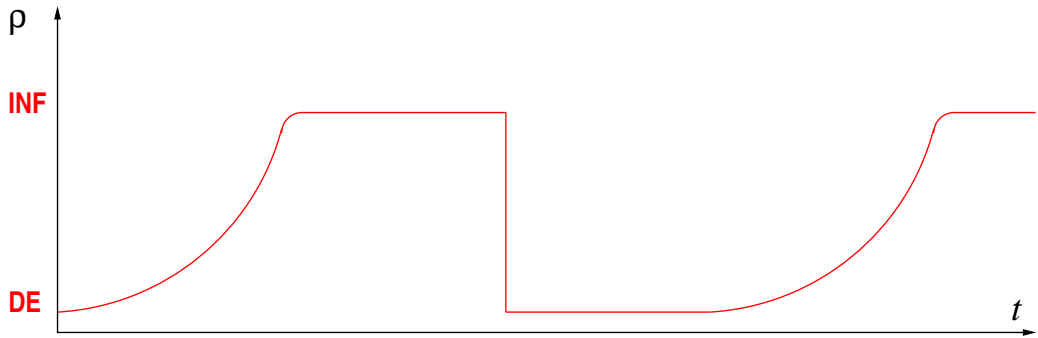}
\caption{%
Schematic evolution of the phantom density $\rho$ in a hypothetical cyclic 
scenario. $\rho$ increases with time approximating a constant $\rho_{\rm max}$, 
which once assumed leads to a period of inflation. Then, some unknown process
reduces drastically the phantom density such that it becomes dark energy, 
until another cycle begins.}
\label{cyclic}
\end{figure}

We can also envisage an interesting hypothetical scenario, where, due to an 
unknown process, the phantom substance drastically decays into a minute residual
density with value \mbox{$\lsim(10^{-3}{\rm eV})^4$}. Its decay products reheat 
the Universe and the hot big bang ensues. Eventually the phantom density takes 
over once more, originally as dark energy,\footnote{Note that the Planck 
satellite observations favour phantom dark energy \cite{planck}.} 
but its density soon increases 
rapidly up to the scale of grand unification $\gsim(10^{16}{\rm GeV})^4$ or so, 
such that in gives rise to another boot of de-Sitter inflation.\footnote{%
Of course, we know that the observed red spectrum of curvature perturbations 
demands that $\dot\rho$ is not zero, but slightly negative during inflation. 
One might hypothesise that the drastic decay of $\rho$ at the end of inflation 
is preceded by some greatly suppressed decay process during inflation. An 
example of this possibility is studied in Ref.~\cite{zhang}.} 
We thus might have a cyclic Universe, which is schematically shown in 
Fig.~\ref{cyclic}.\footnote{This is similar to Ref.~\cite{odintsov2} although
in those works $w$ is taken to vary periodically in time. For a cyclic Universe
due to phantom dark energy in the context of braneworlds see Ref.~\cite{katie}.}
If the scenario can be embedded in string theory \cite{ivonne}, then one might 
wonder if decompactification can happen when the density grows comparable to 
the string scale. Afterwards the extra dimensions might compactify at a 
different Calabi-Yau, implying that the laws of physics might be different
in every cycle. This might lead to a multiverse in time and not in space.


For simplicity, we have assumed isotropy throughout our considerations. In the 
presence of anisotropy the picture could be substantially altered. In many 
cyclic models, for example the Mixmaster universe in Ref~\cite{mixmaster}, 
the cumulative anisotropic effects grow and become overwhelming, possibly 
destabilising the cyclic behaviour. In contrast to the Mixmaster scenario 
however (and other similar cyclic models, e.g. Ref~\cite{other}), the cyclic 
universe considered here does not involve a contracting phase (which would 
enlarge the anisotropy). The total density of the Universe is growing and 
falling but the expansion never halts or reverses itself. Thus, in this case, 
any existing anisotropy might remain negligible, although this needs to be 
investigated.

We have discussed the repercussions of particle production when the Universe is 
dominated by a hypothetical phantom substance. We have argued that backreaction
due to particle production prevents the density from shooting to infinity at a
Big Rip, but instead forces it to stabilise at a large constant value, which 
could be near the string scale or the scale of grand unification, which is a 
little lower. After assuming its constant maximum value there is a period of 
de-Sitter inflation. 
This might lead to a cyclic 
Universe, provided some unknown mechanism terminates inflation and drastically 
reduces the phantom density, giving rise to the thermal bath of the hot big 
bang. 

We have considered Einstein gravity and ignored quantum gravity 
corrections taking the maximum density to remain sub-Planckian.\footnote{For
avoiding the Big Rip singularity in modified gravity see Ref.~\cite{odintsov3}.}
We have not introduced a specific model of phantom dark energy,\footnote{%
We only took \mbox{$w=\,$constant$\,<-1$} for simplicity.} in order to 
emphasise that our treatment and results are generic and due to the existence
of an event horizon in accelerated expansion, which leads to particle production
as in black holes. In that sense, our finding that the Big Rip is unreachable,
seems unavoidable.

\paragraph{Acknowledgements}\leavevmode\\
%
I would like to thank P.~Anderson and G.~Rigopoulos for discussions. My research
is supported (in part) by the Lancaster-Manchester-Sheffield Consortium for 
Fundamental Physics under STFC grant: ST/L000520/1.

\end{document}